\documentclass[journal,draftcls,onecolumn,12pt,twoside]{IEEEtranTCOM}
\normalsize
\usepackage{amsmath}
\usepackage[linesnumbered,ruled,vlined]{algorithm2e}
\usepackage{algorithmic}
\usepackage{graphicx}
\usepackage[cmintegrals]{newtxmath}
\ifCLASSINFOpdf
\else
\fi

\begin{document}
%
\title{Bound on Peak-to-Average Power Ratio with Moment and Reduction Method}
%
%
%

\author{Hirofumi~Tsuda,~\IEEEmembership{Member,~IEEE,}
\thanks{This work was partially supported by JSPS (KAKENHI) Grant Number 18J12903. The material in this paper was presented in part at the ARANFP2019, Marrakech, April 2019 \cite{marrakech}.}
\thanks{The author is with the Department
of Applied Mathematics and Physics, Graduate School of Informatics, Kyoto University, Kyoto,
606-8501, Japan (e-mail: tsuda.hirofumi.38u@st.kyoto-u.ac.jp).}}

%
%

\markboth{IEEE Transactions on Communications}%
{Submitted paper}
%



\maketitle

\begin{abstract}
In this paper, we show bounds on Peak-to-Average Power Ratio (PAPR) and propose a method to reduce PAPR. These bounds are written in terms of moments of transmitted symbols. Further, we show some bounds in special cases and these special bounds include a practical case. From the practical bound, a method to reduce PAPR is derived. Our method is an extension of existing methods, a Partial Transmit Sequence (PTS) technique and a Selective Mapping (SLM) technique. Finally, we numerically verify performance of our method and discuss some conditions for our method. 
\end{abstract}


%
\IEEEpeerreviewmaketitle

\section{Introduction}
%
%
%
%
\IEEEPARstart{O}{rthogonal} Frequency Division Multiplexing (OFDM) systems play important roles in developing communication systems. Signals of OFDM systems consist of sine carves and OFDM systems are implemented by Fast Fourier Transformation (FFT) \cite{ofdmcdma}. It is known that OFDM systems have some advantages. One of them is that OFDM systems can deal with fading effects. This advantage is obtained with a guard interval technique and a zero padding technique \cite{zp}. Since OFDM systems can deal with fading effects, we often assume that a given system is an OFDM system for investigating Multiple Input Multiple Output (MIMO) systems \cite{mimo}. 

However, OFDM systems have some disadvantages. One of them is that OFDM systems have large Peak-to-Average Power Ratio (PAPR) \cite{longlasting}. PAPR is defined as a ratio of the squared maximum amplitude of a given Radio Frequency (RF) signal to its average power \cite{exist} \cite{compute}. Here, large PAPR of a given signal means that the transmitted signal gets more distorted. In ideal cases, an amplifier is ideal: no matter how large an amplitude of a RF signal is, the transmitted signal is not distorted. However, in practical cases, an amplifier has non-linearity and a transmitted signal gets distorted due to the non-linearity of the amplifier if the amplitude of the RF signal is large \cite{concept}. Since distortion noise gets Bit Error Rate (BER) larger, large PAPR should be avoided and lower PAPR has been demanded. Therefore, many methods to reduce PAPR have been proposed \cite{complement}-\cite{timedomain}.

To evaluate PAPR, it is often the case that the Complementary Cumulative Distribution Function (CCDF) of PAPR is evaluated. Since transmitted symbols can be regarded as random variables and PAPR is determined by given transmitted symbols, PAPR can be also regarded as a random variable and the CCDF of PAPR can be considered. Therefore, the CCDF of PAPR is a performance index and we can know the performance of signals if the explicit form of the CCDF of PAPR is known. There are many works to obtain forms of the CCDF of PAPR. If transmitted symbols are identically and independently distributed (iid) random variables and the number of carriers is sufficiently large, then a base-band signal can be regarded as a Gaussian process \cite{convergence}. From this observation, an approximate form of the CCDF of PAPR has been obtained from discrete samples of a base-band signal \cite{asymptotic_papr}. This form is based on independence of each samples. However, it is known that this form underestimates the CCDF \cite{ochiai}. Further, in the Binary Phase Shift Keying (BPSK) scheme, it is proven that the independence of all the sample is not established even if the number of carriers is large \cite{iwasaki}. To overcome this problem, an approximate form based on the distribution on the amplitude of each peak has been proposed \cite{ochiai}. As another example, a form based on the extreme value theory has been proposed \cite{extreme}.

In contrast, if the distributions on transmit symbols are not given, then it is not straightforward to obtain even approximate forms of the CCDF. Thus, in such a case, bounds on the CCDF have been derived instead of obtaining approximate forms \cite{litsyn} \cite{general}. It is often considered to reduce the bound of the CCDF to achieve lower PAPR \cite{general}. In \cite{exist}, the bound on PAPR has been derived and classes of error correction codes achieving low PAPR have been obtained. To obtain upper bounds, some assumptions are often required. One of usual assumptions is about modulation schemes. Thus, with a given modulation scheme, methods to reduce PAPR have been discussed \cite{general} \cite{exist}.

To derive some proposed bounds on the CCDF, the M-QAM scheme and the M-PSK scheme are often assumed. Since a bound obtained under a certain modulation scheme is valid only in the given modulation scheme, the bounds discussed earlier are invalid when a given modulation scheme is neither the M-QAM scheme nor the M-PSK scheme. For example, in the iterative clipping and filtering method \cite{armstrong}, the output symbols may not belong to such a well-known scheme even if the input symbols belong to the M-QAM scheme or the M-PSK scheme. In such a case, some existing bounds cannot be applied. Therefore, to grasp the essential feature of the CCDF of PAPR, it is demanded to obtain a bound under no assumption about a modulation scheme.

In this paper, we derive an upper bound of the CCDF of PAPR with no assumption about a modulation scheme. Here, we make one assumption that the fourth moments of transmitted symbols exist. This assumption is satisfied in practical cases. Then, it turns out that there is a bound written in terms of the fourth moments of transmitted symbols. As a similar bound, it has been proven that there is a bound which is written in terms of moments in BPSK systems \cite{general}. Therefore, our result can be regarded as a generalization of such an existing result. Further, in some cases, special bounds of CCDF are derived. From the derived bounds here, we propose a method to reduce PAPR. This method is based on a technique which has been developed in Independent Component Analysis (ICA) \cite{ica_ori} \cite{ica}. The main idea of ICA is to find a suitable unitary matrix to reduce the kurtosis, which is a statistical quantity written in terms of the fourth moments of signals. From this idea used in ICA, it is expected that our bound can be reduced with unitary matrices since our bound is also written in terms of fourth moments of transmitted symbols. The known methods, a Partial Transmit Sequence (PTS) technique and a Selective Mapping (SLM) method are used to modulate the phase of each symbol so that PAPR decreases \cite{pts} \cite{slm}. Therefore, these known methods are employed to transform codewords with diagonal-unitary matrices and our method can be regarded as a generalization of these methods. The performance of our method is verified numerically. From the numerical results, it is elucidated that PAPR relates to the fourth moments of transmitted symbols. 

\section{OFDM System and PAPR}
In this section, we fix the OFDM system model and the definition of PAPR used throughout this paper. First, a complex baseband OFDM signal is written as \cite{ofdmcdma}
\begin{equation}
  s(t) = \sum_{k=1}^{K} A_k \exp\left(2 \pi j \frac{k-1}{T}t\right), \hspace{2mm} 0 \leq t < T,
  \label{eq:ofdm}
\end{equation}
where $A_k$ is a transmitted symbol, $K$ is the number of symbols, $j$ is the unit imaginary number, and $T$ is a duration of symbols. With Eq. (\ref{eq:ofdm}), a RF OFDM signal is written as

\begin{equation}
  \begin{split}
    \zeta(t) &= \operatorname{Re}\{s(t)\exp(2 \pi j f_c t)\}\\
    &= \operatorname{Re}\left\{\sum_{k=1}^{K} A_k \exp\left(2 \pi j \left(\frac{k-1}{T} + f_c \right)t\right)\right\},
    \end{split}
\label{eq:baseband}
\end{equation}
where $\operatorname{Re}\{z\}$ is the real part of $z$, and $f_c$ is a carrier frequency. With RF signals, PAPR is defined as \cite{exist} \cite{compute}
\begin{equation}
\begin{split}
  & \operatorname{PAPR}(\mathbf{c})\\ 
= & \max_{0 \leq t < T}\frac{\left|\operatorname{Re}\left\{\displaystyle\sum_{k=1}^{K} A_k \exp\left(2 \pi j \left(\frac{k-1}{T} + f_c \right)t\right)\right\}\right|^2}{P_{\operatorname{av}}},
\end{split}
  \label{eq:PAPR}
\end{equation}
where $\mathbf{c} = (A_1,A_2,\ldots,A_K)^\top \in \mathcal{C}$ is called a codeword here, $\mathbf{x}^\top$ is the transpose of $\mathbf{x}$, $\mathcal{C}$ is the set of codewords, $P_{\operatorname{av}}$ corresponds to the average power of baseband-signals, $P_{\operatorname{av}} = \sum_{k=1}^{K}\mathbb{E}[|A_k|^2]$, and $\mathbb{E}[X]$ is the average of $X$. Note that the set of codewords, $\mathcal{C}$, is a subset of $\mathbb{C}^K$, that is, $\mathcal{C} \subseteq \mathbb{C}^K$ and that a codeword is a tuple of transmitted symbols. On the other hand, with baseband signals, Peak-to-Mean Envelope Power Ratio (PMEPR) is defined as \cite{exist} \cite{compute}
\begin{equation}
  \operatorname{PMEPR}(\mathbf{c}) = \max_{0 \leq t < T}\frac{\left|\displaystyle\sum_{k=1}^{K} A_k \exp\left(2 \pi j \frac{k-1}{T} t\right) \right|^2}{P_{\operatorname{av}}}.
  \label{eq:PMEPR}
\end{equation}
As seen in Eqs (\ref{eq:PAPR}) and (\ref{eq:PMEPR}), PAPR and PMEPR are determined by the codeword $\mathbf{c}$ and it is clear that $\operatorname{PAPR}(\mathbf{c}) \leq \operatorname{PMEPR}(\mathbf{c})$ for any codeword $\mathbf{c}$. In \cite{sharif}, the following relation has been proven under the conditions described below
\begin{equation}
  \left( 1 - \frac{\pi^2K^2}{2r^2}\right)\cdot \operatorname{PMEPR}(\mathbf{c}) \leq  \operatorname{PAPR}(\mathbf{c}) \leq  \operatorname{PMEPR}(\mathbf{c}),
  \label{eq:papr_pmepr}
\end{equation}
where $r$ is an integer such that $f_c = r/T$. The conditions that Eq. (\ref{eq:papr_pmepr}) holds are $K \ll r$ and $\exp(2 \pi j K/r) \approx 1$. In addition to these, another relation has been shown in \cite{litsyn}. From Eq. (\ref{eq:papr_pmepr}), PAPR is approximately equivalent to PMEPR for sufficiently large $f_c$. Throughout this paper, we assume that the carrier frequency $f_c$ is sufficiently large, and we consider PMEPR instead of PAPR. Note that this assumption is often used \cite{ochiai}.

\section{Bound on Peak-to-Average Power Ratio}
In this section, we show the bound of the CCDF of PAPR. As seen in Section II, PAPR and PMEPR depend on a given codeword. In analysis of PAPR, it is often the case that a codeword is regarded as a random variable generated from a certain distribution \cite{general}. Then, since a codeword is regarded as a random variable, PAPR and PMEPR are also regarded as random variables. 

First, we make the following assumptions
\begin{itemize}
 \item the probability density of $\mathbf{c}$, $p(\mathbf{c})$, is given and fixed.
 \item the carrier frequency $f_c$ is sufficiently large.
 \item For $1 \leq k,l,m,n \leq K$, the statistical quantity $\mathbb{E}[A_k A_l \overline{A}_m \overline{A}_n]$ exists, where $\overline{z}$ is the conjugate of $z$.
\end{itemize}
The second assumption about a carrier frequency is often used \cite{ochiai}. As seen in Section II, PAPR is approximately equivalent to PMEPR if the carrier frequency $f_c$ is sufficiently large. Thus, we consider PMEPR instead of PAPR. The last assumption has been used in \cite{litsyn} and the quantity $\mathbb{E}[A_k A_l \overline{A}_m \overline{A}_n]$ is called the fourth moment of $A_k$, $A_l$ $\overline{A}_m$ and $\overline{A}_n$. For details about complex multivariate distributions and moments, we refer the reader to \cite{lapidoth} \cite{asymptotic_tech}. From the Cauchy-Schwarz inequality and the last assumption, it follows that
\begin{equation}
 P_{\operatorname{av}} = \sum_{k=1}^K \mathbb{E}[|A_k|^2] \leq \sum_{k=1}^K \sqrt{\mathbb{E}[|A_k|^4]} < \infty.
\end{equation}
Thus, the average power $P_{\operatorname{av}}$ exists, that is, $P_{\operatorname{av}} < \infty$.

Let us consider the PAPR with a given codeword $\mathbf{c}=(A_1, A_2, \ldots, A_K)^\top$. In \cite{tellambura_bound}, the following relation has been proven
\begin{equation}
  \max_t|s(t)|^2 \leq \rho(0) + 2\sum_{k=1}^{K-1}|\rho(k)|,
  \label{eq:bound_env}
\end{equation}
where
\begin{equation}
  \rho(k) = \sum_{l=1}^{K-k}A_l\overline{A}_{k+l}.
\end{equation}
We let $\rho(K)$ be $0$. Note that the quantity $\rho(0)$ is the power of a codeword and that the time $t$ does not appear in the right hand side (r.h.s) of Eq. (\ref{eq:bound_env}). It is not straightforward to analyze Eq. (\ref{eq:bound_env}) since the absolute-value terms appear in Eq. (\ref{eq:bound_env}). To overcome this obstacle, we estimate the upper bound of r.h.s of Eq. (\ref{eq:bound_env}). From the Cauchy-Schwarz inequality, we obtain the following relations
\begin{equation}
\begin{split}
 \max_t|s(t)|^2  & \leq \rho(0) + 2\sum_{k=1}^{K-1}|\rho(k)|\\
 & \leq \sqrt{2K-1}\sqrt{|\rho(0)|^2 + 2\sum_{k=1}^{K-1}|\rho(k)|^2}.
\end{split}
\end{equation}
The above bound is rewritten as
\begin{equation}
 \max_t|s(t)|^4 \leq (2K-1)\left\{|\rho(0)|^2 + 2\sum_{k=1}^{K-1}|\rho(k)|^2\right\}.
\label{eq:bound_cauchy}
\end{equation}
The r.h.s of Eq. (\ref{eq:bound_cauchy}) is rewritten as
\begin{equation}
  \begin{split}
    & (2K-1)\left\{|\rho(0)|^2 + 2\sum_{k=1}^{K-1}|\rho(k)|^2\right\}\\
    =& (2K-1)\left\{\sum_{k=0}^{K-1}|\rho(k)|^2 + \sum_{k=0}^{K-1}|\rho(K-k)|^2 \right\}\\
    =&\frac{2K-1}{2}\left\{\sum_{k=0}^{K-1}|\rho(k) + \overline{\rho(K-k))}|^2\right.\\
    &\left. + \sum_{k=0}^{K-1}|\rho(k) - \overline{\rho(K-k)}|^2\right\}.
  \end{split}
    \label{eq:bound_expansion}
\end{equation}
With the above equations, the r.h.s of  Eq. (\ref{eq:bound_cauchy}) is decomposed into the sum of periodic correlation terms and odd periodic correlation terms. These correlation terms are written as
\begin{equation}
  \begin{split}
    \rho(k) + \overline{\rho(K-k)} &= \mathbf{c}^* B^{(k)}_{1,1}\mathbf{c},\\
    \rho(k) - \overline{\rho(K-k)} &= \mathbf{c}^* B^{(k)}_{-1,1}\mathbf{c},\\
  \end{split}
\end{equation}
where $\mathbf{z}^*$ is the conjugate transpose of $\mathbf{z}$, the matrices $B^{(k)}_{1,1}$ and $B^{(k)}_{-1,1}$ are
\begin{equation}
    B^{(k)}_{1,1} = \left( \begin{array}{cc}
                       O & I_{k}\\
                       I_{K-k} & O
                      \end{array} \right),\hspace{3mm}
   B^{(k)}_{-1,1} = \left( \begin{array}{cc}
                       O & -I_{k}\\
                       I_{K-k} & O
                             \end{array} \right).
\end{equation}
Since these matrices are regular, they can be transformed to diagonal matrices. From this general discussion, these matrices are decomposed with the eigenvalue decomposition as \cite{mypaper} \cite{thesis}
\begin{equation}
  \begin{split}
    B^{(k)}_{1,1} &= V^*D^{(k)}V \qquad B^{(k)}_{-1,1} = \hat{V}^*\hat{D}^{(k)}\hat{V},
  \end{split}
\end{equation}
where $Z^*$ denotes the conjugate transpose of $Z$, $V$ and $\hat{V}$ are unitary matrices whose $(m,n)$-th elements are
\begin{equation}
\begin{split}
V_{m,n} &= \frac{1}{\sqrt{K}}\exp\left(-2 \pi j \frac{mn}{K}\right),\\
 \hat{V}_{m,n} &= \frac{1}{\sqrt{K}}\exp\left(-2 \pi j n\left(\frac{m}{K} + \frac{1}{2K}\right)\right),
\end{split}
\end{equation}
and $D^{(k)}$ and $\hat{D}^{(k)}$ are diagonal matrices whose $n$-th diagonal elements are
\begin{equation}
\begin{split}
D_n^{(k)} &=  \exp\left(-2 \pi j k\frac{n}{K}\right),\\
  \hat{D}_n^{(k)} &=  \exp\left(-2 \pi j k\left(\frac{n}{K} + \frac{1}{2K}\right)\right).
\end{split}
\end{equation}
With these expressions, Eq. (\ref{eq:bound_cauchy}) is written as
\begin{equation}
  \max_t|s(t)|^4 \leq \frac{K(2K-1)}{2}\left\{\sum_{k=1}^{K}|\alpha_k|^4 + \sum_{k=1}^{K}|\beta_k|^4\right\},
\end{equation}
where $\alpha_k$ and $\beta_k$ are the $k$-th element of $\boldsymbol{\alpha}$ and $\boldsymbol{\beta}$ written as $\boldsymbol{\alpha}=V\mathbf{c}$ and $\boldsymbol{\beta} = \hat{V}\mathbf{c}$, respectively. With the codeword $\mathbf{c}$, the above inequality is written as
\begin{equation}
\begin{split}
 &  \max_t|s(t)|^4 \\
\leq & \frac{K(2K-1)}{2}\sum_{k=1}^K \left\{\left(\mathbf{c}^*V^*G_k V \mathbf{c}\right)^2 + \left(\mathbf{c}^*
      \hat{V}^* G_k \hat{V} \mathbf{c}\right)^2\right\},
\end{split}
  \label{eq:bound_expansion2}
\end{equation}
where $G_k$ is a matrix whose $(k,k)$-th element is unity and the other elements are zero. Note that $G_k^*G_k = G_k$. For the later convenience, we set $C_k = V^*G_k V$ and $\hat{C}_k = \hat{V}^*G_k \hat{V}$, respectively. Note that the matrices $C_k$ and $\hat{C}_k$ are positive semidefinite Hermitian matrices since $C_k$ and $\hat{C}_k$ are the Gram matrices. From Eq. (\ref{eq:bound_expansion2}), with a given codeword $\mathbf{c}$, the bound of the squared PAPR is obtained as

\begin{equation}
\begin{split}
 \operatorname{PAPR}(\mathbf{c})^2 &\leq \frac{\max_t|s(t)|^4}{P_{\operatorname{av}}^2}\\
&\leq \frac{K(2K-1)}{2P_{\operatorname{av}}^2}\sum_{k=1}^K \left\{\left(\mathbf{c}^* C_k \mathbf{c}\right)^2 + \left(\mathbf{c}^*
     \hat{C}_k \mathbf{c}\right)^2\right\}.
\end{split}
\end{equation}
In the above relations, the first inequality is obtained from the result that $\operatorname{PAPR}(\mathbf{c}) \leq \operatorname{PMEPR}(\mathbf{c})$.

From the above discussions, we have arrived at the bound on PAPR with a given codeword $\mathbf{c}$. From this bound, we can obtain the bound of the CCDF of PAPR as follows. Let $\operatorname{Pr}(\operatorname{PAPR} > \gamma)$ be the CCDF of PAPR, where $\gamma$ is a positive real number. Then, the following relations are obtained \cite{marrakech}
 \begin{equation}
  \begin{split}
  &\operatorname{Pr}(\operatorname{PAPR} > \gamma)\\
 = &\operatorname{Pr}(\operatorname{PAPR}^2 > \gamma^2)\\ 
\leq & \operatorname{Pr}(\max_t|s(t)|^4 > P_{\operatorname{av}}^2\gamma^2)\\
\leq & \frac{\mathbb{E}\left[\max_t|s(t)|^4 \right]}{P_{\operatorname{av}}^2\gamma^2}\\
\leq & \frac{K(2K-1)}{2P_{\operatorname{av}}^2 \gamma^2} \sum_{k=1}^K \mathbb{E}\left[\left(\mathbf{c}^* C_k \mathbf{c}\right)^2 + \left(\mathbf{c}^* \hat{C}_k \mathbf{c}\right)^2\right].
  \end{split}
\label{eq:prob_bound}
 \end{equation}
In the course of deriving Eq. (\ref{eq:prob_bound}), the first equation has been obtained from the fact that PAPR is positive. The first inequality has been obtained from Eq. (\ref{eq:PMEPR}) and the fact that $\operatorname{PAPR}(\mathbf{c}) \leq \operatorname{PMEPR}(\mathbf{c})$ for any codeword $\mathbf{c}$. The second inequality has been obtained with the Markov inequality \cite{prob}. The last inequality has been obtained from Eq. (\ref{eq:bound_expansion2}). 

As seen in Eq. (\ref{eq:prob_bound}), the bound of the CCDF is written in terms of the fourth moments of codewords and this bound does not depend on a modulation scheme. 

\section{Bounds in Special Cases} 
In the previous section, we have seen the bound of the CCDF written in terms of the fourth moments of codewords. To derive this bound, we have made one assumption about codewords that their fourth moments exist. In this section, we show bounds with two special cases: one is the case where codewords are generated from the Gaussian distribution and the other is the case where there is a norm condition of codewords. 

\subsection{Bound with Codewords Generated from Gaussian Distribution}
First, we consider the case where codewords are generated from the Gaussian distribution whose mean is the zero vector. In information theory, we often consider input symbols which follow the Gaussian distribution since the Gaussian input achieves largest information capacity in Gaussian channel \cite{cover}. 

Here, we assume that codewords are generated from the complex multivariate Gaussian distribution
\begin{equation}
 \mathbf{c} \sim \mathcal{CN}(\mathbf{0},\Sigma),
\end{equation}
where $\mathcal{CN}(\boldsymbol{\mu},\Sigma)$ denotes the complex multivariate Gaussian distribution whose mean and covariance matrix are $\boldsymbol\mu$ and $\Sigma$, respectively. The definition and properties of the multivariate Gaussian distribution have been shown in \cite{graphical}. Then, from Eq. (\ref{eq:prob_bound}), the bound of the CCDF is written as
 \begin{equation}
  \begin{split}
  &\operatorname{Pr}(\operatorname{PAPR} > \gamma)\\
\leq & \frac{3K(2K-1)}{2P_{\operatorname{av}}^2 \gamma^2} \sum_{k=1}^K \left\{\operatorname{Tr}\left(C_k \Sigma\right)^2 + \operatorname{Tr}\left(\hat{C}_k \Sigma\right)^2\right\},
  \end{split}
\label{eq:bound_gauss}
 \end{equation}
where $\operatorname{Tr}(X)$ is the trace of $X$. The proof is written in Appendix A. From Eq. (\ref{eq:bound_gauss}), the bound of the CCDF with codewords generated from the Gaussian distribution is written in terms of its covariance matrix. Note that the above bound is also valid for codewords generated from the real-valued multivariate Gaussian distribution whose mean and covariance matrix are zero and $\Sigma$, respectively.

\subsection{Bounds with Codewords Generated from Practical Scheme}
We have derived the bound of CCDF with Gaussian inputs. Here, we consider a practical case, that is, the support of the probabilistic density function of codewords lies in a certain compact set.

To derive the bound, we make the following assumption
\begin{equation}
 \max_{\mathbf{c} \in \mathcal{C}} \|\mathbf{c}\|_{l_1} < \infty,
\label{eq:norm_assumption}
\end{equation} 
where $\|\mathbf{z}\|_{l_1}$ denotes the $l_1$ norm of $\mathbf{z}$. Thus, the support of the probability density function of codewords lies in a compact set $S_r = \{\mathbf{c} \mid \|\mathbf{c}\|_{l_1} \leq r\}$, where $r =  \max_{\mathbf{c} \in \mathcal{C}} \|\mathbf{c}\|_{l_1}$. In practical cases, for example, a M-QAM scheme and a M-PSK scheme, the above condition is clearly satisfied. From this condition, the maximum amplitude of a base-band signal for a certain codeword has the following bounds
\begin{equation}
\begin{split}
\|\mathbf{c}\|^2_{l_2} \leq  |s(t)|^2 &\leq \|\mathbf{c}\|^2_{l_1} \leq K \|\mathbf{c}\|^2_{l_2},
\end{split} 
\label{eq:papr_bounds}
\end{equation}
where $\|\mathbf{z}\|_{l_2}$ denotes the $l_2$ norm of $\mathbf{z}$. The proof goes as follows. First, from the H\"older's inequality,
\begin{equation}
 \begin{split}
\|\mathbf{c}\|^2_{l_2} &= \frac{1}{T}\int_0^T \left|\sum_{k=1}^K A_k \exp\left(2 \pi j \frac{k-1}{T}t\right)\right|^2 dt\\
& \leq \max_{t} \left|\sum_{k=1}^K A_k \exp\left(2 \pi j \frac{k-1}{T}t\right)\right|^2.
 \end{split}
\end{equation}
In the above inequality, we have used the property that OFDM baseband signals have a periodic function with the duration $T$. The upper bounds of the maximum amplitude are proven by the triangle inequality and the Cauchy-Schwarz inequality.

From Eqs. (\ref{eq:norm_assumption}) and (\ref{eq:papr_bounds}), there exist two numbers such that
\begin{equation}
\begin{split}
 a &= \min_{\mathbf{c} \in \mathcal{C}} |s(t)|^2\\
 b &= \max_{\mathbf{c} \in \mathcal{C}} |s(t)|^2.
\end{split}
\end{equation}
Then, from Eqs. (\ref{eq:PMEPR}) and (\ref{eq:papr_bounds}), the support of PMEPR is compact, that is, it follows
\begin{equation}
 \operatorname{Pr}\left(\frac{a}{P_{\operatorname{av}}} \leq \operatorname{PMEPR} \leq \frac{b}{P_{\operatorname{av}}}\right) = 1.
\label{eq:hoeffding_cond}
\end{equation}
With these $a$ and $b$, for sufficient large $\gamma$, there is the following bound of the CCDF
 \begin{equation}
  \begin{split}
 &\operatorname{Pr}(\operatorname{PAPR} > \gamma)\\
\leq & \exp\left(-\frac{2}{(b^2-a^2)}\left(P^2_{\operatorname{av}}\gamma^2 - R\right)^2\right),
  \end{split}
\label{eq:bound_hoeffding}
 \end{equation}
where
\begin{equation}
R = \frac{K(2K-1)}{2}\sum_{k=1}^K \mathbb{E}\left[\left(\mathbf{c}^* C_k \mathbf{c}\right)^2 + \left(\mathbf{c}^* \hat{C}_k \mathbf{c}\right)^2\right].
\end{equation}
The proof is written in Appendix B. As seen in Appendix B, there is a condition that $\gamma^2 > R/P_{\operatorname{av}}^2$. This bound is tighter than the bound shown in Eq. (\ref{eq:prob_bound}) in a sense of the order of $\gamma$.

From Eq. (\ref{eq:bound_hoeffding}), we can derive the bound of the CCDF in the M-QAM scheme. First, the set of codewords is defined as \cite{general}
\begin{equation}
\begin{split}
 \mathcal{C} =& \{\mathbf{c} \mid A_i \in D((2m_1-1)+j(2m_2-1)),\\
& m_1,m_2 \in \{-m/2+1,\ldots,m/2\}\},
\end{split}
\end{equation}
where $m>1$, $M$, and $D$ are the numbers such that $M=m^2$ and $D^2=3/2(M-1)$. Then, the numbers $a$ and $b$ can be chosen as
\begin{equation}
 a = 0,\qquad b = 2K^2 D^2(\sqrt{M}-1)^2.
\end{equation}
It is clear that these $a$ and $b$ satisfy Eq. (\ref{eq:hoeffding_cond}). Thus, we have
 \begin{equation}
  \begin{split}
 &\operatorname{Pr}(\operatorname{PAPR} > \gamma)\\
\leq & \exp\left(-\frac{1}{2K^4D^4(\sqrt{M}-1)^4}\left(P^2_{\operatorname{av}}\gamma^2 - R\right)^2\right).
  \end{split}
 \end{equation}

\section{PAPR Reduction with Unitary Matrix}
In the previous section, we have derived the bounds of the CCDF in some special cases. In particular, in the case where the support of the probability distribution of codewords lies in a compact set, there is a tighter bound than one derived in Section III in a sense of the order of $\gamma$. In this section, we derive a method to reduce PAPR from the bound derived in the previous section. This method is based on the Independent Component Analysis (ICA) technique since we have seen that PAPR relates to the fourth moments of codewords and the aim of ICA technique is also to reduce the fourth moments of signals. Thus, we can apply the methods in the ICA technique to the PAPR-reduction.

To derive the method to reduce PAPR, we make the following assumptions
\begin{itemize}
 \item the number of components in codewords, $|C|$, is finite, that is, $|\mathcal{C}| <\infty$.
\item each codeword is chosen with equal probability from $\mathcal{C}$.
\item $ \max_{\mathbf{c} \in \mathcal{C}} \|\mathbf{c}\|_{l_1} < \infty$.
\end{itemize}
From the above assumptions and Eq. (\ref{eq:bound_hoeffding}), the bound of the CCDF is rewritten as
 \begin{equation}
  \begin{split}
 &\operatorname{Pr}(\operatorname{PAPR} > \gamma)\\
\leq & \exp\left(-\frac{2}{(b^2-a^2)}\left(P^2_{\operatorname{av}}\gamma^2 - R(\mathcal{C})\right)^2\right),
  \end{split}
\label{eq:bound_hoeffding_sample}
 \end{equation}
where
\begin{equation}
R(\mathcal{C}) = \frac{K(2K-1)}{2|\mathcal{C}|}\sum_{\mathbf{c} \in \mathcal{C}}\sum_{k=1}^K \left\{\left(\mathbf{c}^* C_k \mathbf{c}\right)^2 + \left(\mathbf{c}^* \hat{C}_k \mathbf{c}\right)^2\right\}.
\end{equation} 
Here, the numbers $a$ and $b$ are chosen as
\begin{equation}
 a = \min_{\mathbf{c} \in \mathcal{C}}\|\mathbf{c}\|^2_{l_2}, \qquad  b = K \max_{\mathbf{c} \in \mathcal{C}}\|\mathbf{c}\|^2_{l_2}.
\end{equation}
Then, Eq. (\ref{eq:hoeffding_cond}) is satisfied. As seen in Eq. (\ref{eq:bound_hoeffding_sample}), the bound is written in terms of the fourth moments of codewords. Thus, it is expected that PAPR reduces as the fourth moments of codewords gets smaller. In the ICA technique, an unitary matrix is used to reduce the fourth moments. In this section, our goal is to find unitary matrices which achieve low PAPR. 

From the assumptions, since the number of codewords is finite, we can decompose the set of codewords into $N$ subsets such that
\begin{equation}
 \mathcal{C} = \bigcup_{n=1}^N \mathcal{C}_n, \hspace{2mm} \mathcal{C}_m \cap \mathcal{C}_n = \emptyset \hspace{2mm}\mbox{for} \hspace{2mm}m \neq n. 
\end{equation}
Let $W_n$ be a unitary matrix for $n=1,\ldots,N$ and these matrices are used to reduce PAPR. The scheme of our method is described as follows. First, let the transmitter and the receiver know the unitary matrices $\{W_n\}_{n=1}^N$. At the transmitter side, each codeword $\mathbf{c} \in \mathcal{C}_i$ is modulated to $W_i \mathbf{c}$ with the unitary matrix $W_i$. Then, the transmitter sends the number $i$ and $W_i \mathbf{c}$. At the receiver side, the symbol $\mathbf{y}$ and the number $i$ are received. Then, the receiver estimates the codeword $\hat{\mathbf{c}}$ as $\hat{\mathbf{c}} = W^*_i \mathbf{y}$. It is clear that $\hat{\mathbf{c}} = \mathbf{c}$ if $\mathbf{y} = W_i\mathbf{c}$. In the above scheme, the bound in Eq. (\ref{eq:bound_hoeffding_sample}) is rewritten as
 \begin{equation}
  \begin{split}
 &\operatorname{Pr}(\operatorname{PAPR} > \gamma)\\
\leq & \exp\left(-\frac{2}{(b^2-a^2)}\left(P^2_{\operatorname{av}}\gamma^2 - R(\{W_n\mathcal{C}_n\}_{n=1}^N)\right)^2\right),
  \end{split}
\label{eq:bound_hoeffding_unitary}
 \end{equation}
where
\begin{equation}
\begin{split}
&R(\{W_n\mathcal{C}_n\}_{n=1}^N)= \frac{K(2K-1)}{2|\mathcal{C}|}\sum_{n=1}^N\sum_{\mathbf{c} \in \mathcal{C}_n}\sum_{k=1}^K \\
& \left\{\left(\mathbf{c}^*W_n^* C_k W_n\mathbf{c}\right)^2 + \left(\mathbf{c}^*W_n^* \hat{C}_k W_n\mathbf{c}\right)^2\right\}.
\end{split}
\end{equation} 
Note that the numbers $a$ and $b$ can be regarded as constants since an unitary matrix preserves the $l_2$ norm, that is, for any $\mathbf{z} \in \mathbb{C}^n$, $\|\mathbf{z}\|_{l_2} = \|W\mathbf{z}\|_{l_2}$, where $W$ is a $n \times n$ unitary matrix. 

Let us consider the case where the channel is a Gaussian channel and the codeword $\mathbf{c} \in \mathcal{C}_i$ is sent. In such a situation, the received symbol $\mathbf{y}$ is written as 
\begin{equation}
 \mathbf{y} = W_i \mathbf{c} + \mathbf{n},
\end{equation}
where $\mathbf{n}$ is a noise vector whose components follow the complex Gaussian distribution independently. Then, the estimated codeword is written as
\begin{equation}
 \hat{\mathbf{c}} = \mathbf{c} + W_i^*\mathbf{n}.
\end{equation}
From the above equation, SNR is preserved through our method since the matrix $W_i$ is unitary. 

In existing methods, a PTS technique and a SLM method, one diagonal unitary matrix corresponds to one codeword. By contrast, in our method, one unitary matrix corresponds to one set of codewords. This is the main difference between our method and the existing methods.  

We have proposed a scheme to reduce PAPR. The remained problem is to find $W_n$ which reduces the bound and achieves low PAPR for $n=1,\ldots,N$. To find such matrices, we consider the gradient method. For a fixed $\gamma$, the r. h. s of Eq. (\ref{eq:bound_hoeffding_unitary}) is a function with respect to $W_n$. To apply the gradient method, we have to calculate the derivative of the function with respect to $W_n$. However, since expressions involving a complex conjugate or a conjugate transpose do not satisfy the Cauchy-Riemann equations in general \cite{cookbook}, the function is not differentiable. To overcome this obstacle, we introduce another differentiation called the generalized complex gradient \cite{complex_gradient}
\begin{equation}
 \frac{\partial F}{\partial W_n} = \frac{\partial F}{\partial \operatorname{Re}\{W_n\}} + j\frac{\partial F}{\partial \operatorname{Im}\{W_n\}},
\end{equation}
where $F$ is a function with respect to the variable $W_n$, $\operatorname{Re}\{Z\}$ and $\operatorname{Im}\{Z\}$ are the real part and imaginary part of the matrix $Z$, respectively. With this definition, we can obtain the differentiation. Let us define $f$ as
\begin{equation}
 f = \exp\left(-\frac{2}{(b^2-a^2)}\left(P^2_{\operatorname{av}}\gamma^2 - R(\{W_n\mathcal{C}_n\}_{n=1}^N)\right)^2\right).
\end{equation} 
This function is the r. h. s of the Eq. (\ref{eq:bound_hoeffding_unitary}). From the above definition, we have
\begin{equation}
\begin{split}
  &\frac{\partial f}{\partial W_n}= \epsilon(\gamma,\{W_n\})\\
& \cdot \sum_{\mathbf{c} \in \mathcal{C}_n} \sum_{k=1}^K \left\{\left(\mathbf{c}^* W^*_n C_k W_n \mathbf{c}\right)C_k  + \left(\mathbf{c}^* W^*_n \hat{C}_k W_n \mathbf{c}\right)\hat{C}_k \right\}W_n \mathbf{c} \mathbf{c}^*,
\end{split}
\label{eq:gradient}
\end{equation}
where
\begin{equation}
\begin{split}
 \epsilon(\gamma,\{W_n\}) =& \frac{8K(2K-1)}{|\mathcal{C}|(b^2 -a^2)}\left(P^2_{\operatorname{av}}\gamma^2 - R(\{W_n\mathcal{C}_n\}_{n=1}^N)\right) \\
& \cdot\exp\left(-\frac{2}{(b^2-a^2)}\left(P^2_{\operatorname{av}}\gamma^2 - R(\{W_n\mathcal{C}_n\}_{n=1}^N)\right)^2\right).
\end{split}
\end{equation}
If $\gamma^2 > R(\{W_n\mathcal{C}\}_{n=1}^N)/P_{\operatorname{av}}^2$, then the bound is valid and the scalar $\epsilon(\gamma,\{W_n\})$ is positive. In the gradient method, the update rule at the $l$-th iteration is written as 
\begin{equation}
  W^{(l+1)}_n \leftarrow W^{(l)}_n - \hat{\epsilon} \frac{\partial f}{\partial W^{(l)}_n},
\label{eq:gradient_method}
\end{equation}
where $W^{(l)}_n$ is the matrix obtained at the $l$-th iteration and $\hat{\epsilon}$ denotes a step size. Here, $\hat{\epsilon}$ is sufficiently small and its reason is given later. Since the gradient in Eq. (\ref{eq:gradient}) is decomposed into the scalar $\epsilon(\gamma,\{W_n\})$ and the matrix, for a certain $\gamma$, Eq. (\ref{eq:gradient_method}) is rewritten as
\begin{equation}
  W^{(l+1)}_n \leftarrow W^{(l)}_n - \epsilon \cdot \Delta W^{(l)}_n,
\label{eq:update_rule}
\end{equation}
where
\begin{equation}
\begin{split}
 \Delta W^{(l)}_n =& \sum_{\mathbf{c} \in \mathcal{C}_n} \sum_{k=1}^K \left\{\left(\mathbf{c}^* \left(W^{(l)}_n\right)^* C_k W^{(l)}_n \mathbf{c}\right)C_k  \right. \\
&+ \left. \left(\mathbf{c}^* \left(W^{(l)}_n\right)^* \hat{C}_k W^{(l)}_n \mathbf{c}\right)\hat{C}_k \right\} W^{(l)}_n \mathbf{c} \mathbf{c}^* 
\end{split}
\label{eq:delta_w}
\end{equation}
and $\epsilon = \hat{\epsilon} \cdot \epsilon(\gamma,\{W_n\})$. If $\hat{\epsilon}$ is chosen to be sufficiently small, then $\epsilon$ can be small.

From Eq. (\ref{eq:update_rule}), we obtain the updated matrices $W^{(l+1)}_n$ for $n=1,\ldots,N$.  However, the matrices $W_n^{(l+1)}$ may not be unitary matrices. Thus, these matrices have to be projected onto the region of unitary matrices. One method is to use the Gram-Schmidt process \cite{ica} \cite{ica_app} \cite{ica_complex}. First, we decompose the matrix $W^{(l+1)}_n$ as $W^{(l+1)}_n = (\mathbf{w}_1,\ldots,\mathbf{w}_K)^\top$ and update $\mathbf{w}_1 \leftarrow \mathbf{w}_1/\|\mathbf{w}_1\|_2$. Then, the following steps are iterated for $k=2,\ldots,K$
\begin{enumerate}
 \item $\mathbf{w}_k \leftarrow \mathbf{w}_k - \sum_{i=1}^{k-1}\mathbf{w}^*_k \mathbf{w}_i \mathbf{w}_i$.
\item $\mathbf{w}_k \leftarrow \mathbf{w}_k/\|\mathbf{w}_k\|_2$.
\end{enumerate}
Finally, the projected matrix is obtained as $W^{(l+1)}_n = (\mathbf{w}_1,\ldots,\mathbf{w}_K)^\top$.

With the Gram-Schmidt process, we can obtain the unitary matrices. However, it is unclear what the suitable order to choose and to normalize vectors is. To avoid this ambiguity, a symmetric decorrelation technique has been proposed \cite{ica_app} \cite{ica} \cite{ica_complex}. A symmetric decorrelation technique is to normalize $W^{(l+1)}_i$ as
\begin{equation}
 W^{(l+1)}_n \leftarrow \left(W^{(l+1)}_n \left(W^{(l+1)}_n\right)^*\right)^{-1/2}W^{(l+1)}_n,
\label{eq:proj_unitary}
\end{equation}
where $(ZZ^*)^{-1/2}$ is obtained from the eigenvalue decomposition of $ZZ^* = F \Lambda F^*$ as $F\Lambda^{-1/2}F^*$ with $F$ being a unitary matrix, $\Lambda$ and $\Lambda^{-1/2}$ being diagonal positive matrices written as $\Lambda = \operatorname{diag}(\lambda_1,\lambda_2,\ldots,\lambda_K)$ and $\Lambda^{-1/2} = \operatorname{diag}(\lambda_1^{-1/2},\lambda_2^{-1/2},\ldots,\lambda_K^{-1/2})$, respectively. With the above projection, we can obtain the unitary matrix $W^{(l+1)}_n$. The algorithm of our method is summarized in Algorithm \ref{algo:unitary_method}.

\begin{algorithm}[htbp]
Set $\epsilon$, the initial unitary matrix $W^{(1)}_n$ for $n=1,2,\ldots,N$ and the iteration count $l=1$.\\
For $n=1,2,\ldots,N$, calculate $\Delta W^{(l)}_n$ and obtain the matrix $W^{(l+1)}_n$ as
\begin{equation*}
 W^{(l+1)}_n \leftarrow W^{(l)}_n - \epsilon \Delta W^{(l)}_n.
\end{equation*} \\
Project $W^{(l+1)}_n$ onto the set of unitary matrices for $n=1,2,\ldots,N$ with Gram-Schmidt process or Eq. (\ref{eq:update_rule}).\\
Let $\|W\|$ be the norm of the matrix $W$. If $\|W^{(l+1)}_n - W^{(l)}_n\| \approx 0$ for $n=1,2,\ldots,N$ or the iteration count $l$ gets sufficiently large, then stop. Otherwise, set $l \leftarrow l+1$ and go to step 2.
 \caption{How to Find Unitary Matrix in Our Method}
 \label{algo:unitary_method}
\end{algorithm}
In Algorithm \ref{algo:unitary_method}, the matrix $W^{(l+1)}_n$ obtained at step 2 has to be full rank. Since the matrix $W^{(l)}_n$ is an unitary matrix, the matrix $W^{(l+1)}_n$ is full-rank with sufficiently small $\epsilon$. This is the reason why $\epsilon$ has to be sufficiently small.

\section{Stochastic Gradient Method}
In the previous section, we have proposed the method to reduce PAPR with unitary matrices and how to find the matrices. To find the matrices, the gradient method is used. In Eq. (\ref{eq:update_rule}), to calculate the matrix $\Delta W^{(l)}_n$, $|\mathcal{C}_n|$ matrices have to be calculated since the differentiation consists of the sum of the matrices for codewords belonging to $\mathcal{C}_n$. This implies that the larger calculation amount may be required as the number of codewords in $\mathcal{C}_n$ gets larger. To avoid this, we apply the stochastic gradient method \cite{stochastic_grad} \cite{stochastic_grad2} to Eq. (\ref{eq:update_rule}). 

Here, for $n=1,\ldots,N$, we choose a certain $\mathbf{c} \in \mathcal{C}_n$ with the probability $\operatorname{Pr}(\mathbf{c}) = 1/|\mathcal{C}_n|$. With this $\mathbf{c}$, we calculate the following quantity
\begin{equation}
\begin{split}
 \Delta W^{(l)}_{n,\mathbf{c}} =& \sum_{k=1}^K \left\{\left(\mathbf{c}^* \left(W^{(l)}_n\right)^* C_k W^{(l)}_n \mathbf{c}\right)C_k  \right. \\
&+ \left. \left(\mathbf{c}^* \left(W^{(l)}_n\right)^* \hat{C}_k W^{(l)}_n \mathbf{c}\right)\hat{C}_k \right\} W^{(l)}_n \mathbf{c} \mathbf{c}^*.
\end{split}
\end{equation}
This $\Delta W^{(l)}_{n,\mathbf{c}}$ appeared in Eq. (\ref{eq:delta_w}). With this $\Delta W^{(l)}_{n,\mathbf{c}}$, we obtain $W^{(l+1)}_n$. The algorithm is written in Algorithm \ref{algo:unitary_method_stoc}. Note that the feasible region in this algorithm is a set of unitary matrices and not convex.

\begin{algorithm}[htbp]
Set $\epsilon$, the initial unitary matrix $W^{(1)}_n$ for $n=1,2,\ldots,N$ and the iteration count $l=1$.\\
For $n=1,2,\ldots,N$, choose $\mathbf{c} \in \mathcal{C}_n$ with the probability $\operatorname{Pr}(\mathbf{c})=1/|\mathcal{C}_n|$. Then, calculate $\Delta W^{(l)}_{n,\mathbf{c}}$ and obtain the matrix $W^{(l+1)}_n$ as
\begin{equation*}
 W^{(l+1)}_n \leftarrow W^{(l)}_n - \epsilon \Delta W^{(l)}_{n,\mathbf{c}}.
\end{equation*} \\
Project $W^{(l+1)}_n$ onto the set of unitary matrices for $n=1,2,\ldots,N$ with Gram-Schmidt process or Eq. (\ref{eq:update_rule}).\\
Let $\|W\|$ be the norm of the matrix $W$. If $\|W^{(l+1)}_n - W^{(l)}_n\| \approx 0$ for $n=1,2,\ldots,N$ or the iteration count $l$ gets sufficiently large, then stop. Otherwise, set $l \leftarrow l+1$ and go to step 2.
 \caption{How to Find Unitary Matrix with Stochastic Gradient Method}
 \label{algo:unitary_method_stoc}
\end{algorithm}

%

\section{Numerical Results}
In this Section, we show the performance of our proposed method. As seen in Section II, we assume that the carrier frequency $f_c$ in Eqs (\ref{eq:baseband}) and (\ref{eq:PAPR}) is sufficiently large and then PAPR is approximately equivalent to PMEPR. Thus, we measure PMEPR instead of PAPR.  We set the parameters as $K=128$ and $|\mathcal{C}|=2000$. To measure PMEPR, we choose oversampling parameter $J=16$. How to choose the parameter $J$ has been discussed in \cite{sharif} \cite{clip}. The modulation scheme is 16-QAM. All symbols are generated independently from the 16-QAM set and then we obtain the set of codewords $\mathcal{C} = \left\{ \mathbf{c}_1,\ldots,\mathbf{c}_{|\mathcal{C}|} \right\}$. In each $N$, we set the subsets of codewords as
\begin{equation}
 \mathcal{C}_n = \left\{ \mathbf{c}_{\frac{|\mathcal{C}|}{N}(n-1)+1}, \mathbf{c}_{\frac{|\mathcal{C}|}{N}(n-1)+2}, \ldots, \mathbf
{c}_{\frac{|\mathcal{C}|n}{N}} \right\}
\end{equation}
for $n=1,2,\ldots,N$. Thus, each subset of codewords is randomly obtained from the original 16-QAM set and the number of components in $\mathcal{C}_n$ is $|\mathcal{C}|/N$ for $n=1,2,\ldots,N$.
As initial points, we set $W_n^{(1)} = E$, where $E$ is identity matrix for $n=1,2,\ldots,N$. The gradient parameter $\epsilon$ is set as $\epsilon = K^{-3/2}$. To find unitary matrices, we have used the Algorithm \ref{algo:unitary_method_stoc} and have used the symmetric decorrelation technique described in Eq. (\ref{eq:proj_unitary}).

\begin{figure}[htbp]
\centering  
\includegraphics[width=3.0in]{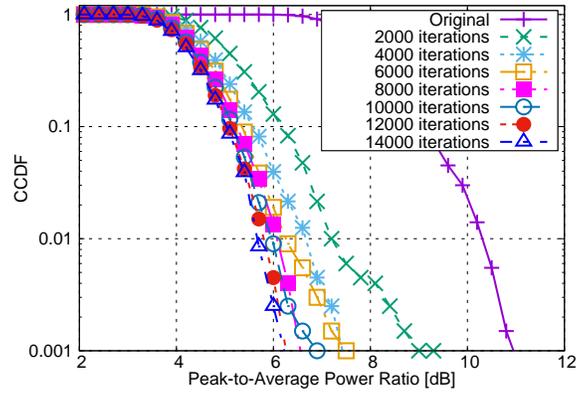}
\caption{PAPR in each iteration with $N=5$}
\label{fig:PAPR_iter_N5}
\end{figure}

Figure \ref{fig:PAPR_iter_N5} shows PAPR in our method with the parameter $N=5$. Each curve in the figure corresponds to the number of the iteration steps. As seen in this figure, PAPR gets smaller as the number of the iteration steps increases and PAPR seems to converge to one with 14000 iterations. This result shows that we can obtain the unitary matrices which achieve lower PAPR as the number of the iteration steps increases. Since our method is to reduce the bound on PAPR in Eq. (\ref{eq:bound_hoeffding_unitary}), this result implies that decreasing our bound may lead to decreasing PAPR. We conclude that our bound closely relates to the CCDF of PAPR. 

\begin{figure}[htbp]
\centering  
\includegraphics[width=3.0in]{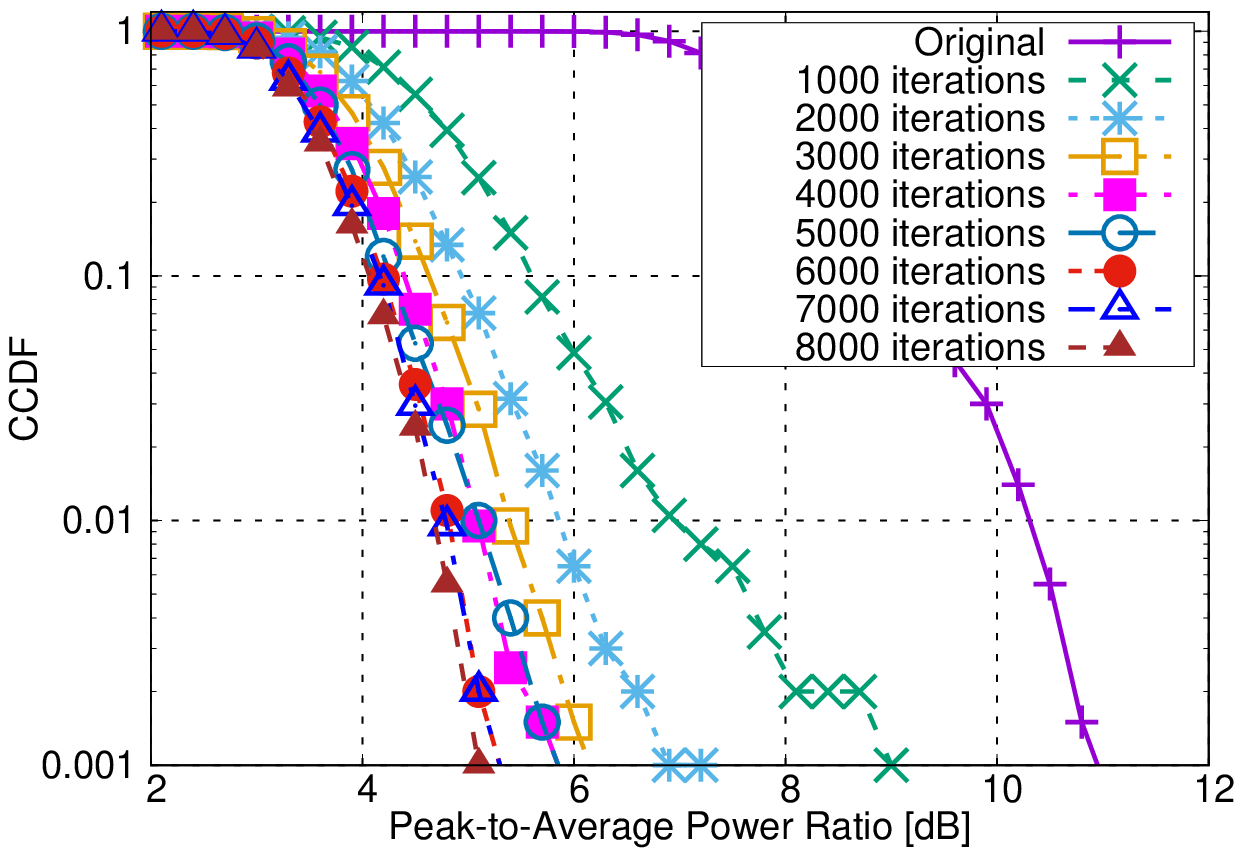}
\caption{PAPR in each iteration with $N=10$}
\label{fig:PAPR_iter_N10}
\end{figure}

Figure \ref{fig:PAPR_iter_N10} shows PAPR in our method with the parameter $N=10$. Similar to the result with the parameter $N=5$, our method achieves lower PAPR when the number of the iteration steps gets larger and PAPR seems to converge to one with 8000 iterations. However, from Figs. \ref{fig:PAPR_iter_N5} and \ref{fig:PAPR_iter_N10}, the converged PAPR with $N=10$ is lower than one with $N=5$. The reason may be explained as follows. First, we define the following quantity
\begin{equation}
g(\mathcal{C}_n) = \frac{1}{|\mathcal{C}_n|}\sum_{\mathbf{c} \in \mathcal{C}_n} \mathbf{c}\mathbf{c}^*.
\end{equation}
If the mean of $\mathbf{c} \in \mathcal{C}_n$ is zero, then the above quantity $g(\mathcal{C}_n)$ is a covariance matrix. From Eq. (\ref{eq:bound_hoeffding_unitary}), the bound depends on the unitary matrices $\{W_n\}_{n=1}^N$ through the quantity $R(\{W_n\mathcal{C}_n\})_{n=1}^N$. Thus, we consider only the quantity $R(\{W_n\mathcal{C}_n\})_{n=1}^N$. From the Jensen's inequality \cite{billingsley}, if each $g(\mathcal{C}_n)$ equals identity matrix, then the quantity $R(\{W_n\mathcal{C}_n\})_{n=1}^N$ has the following bound
\begin{equation}
\begin{split}
&R(\{W_n\mathcal{C}_n\})_{n=1}^N \\
= & \frac{K(2K-1)}{2|\mathcal{C}|} \sum_{n=1}^N \sum_{\mathbf{c} \in \mathcal{C}_n}\sum_{k=1}^K \left(\mathbf{c}^* W_n^* C_k W_n\mathbf{c}\right)^2 + \left(\mathbf{c}^* W_n^* \hat{C}_k W_n \mathbf{c}\right)^2 \\
= & \frac{K(2K-1)}{2} \sum_{n=1}^N \frac{|\mathcal{C}_n|}{|\mathcal{C}|} \frac{1}{|\mathcal{C}_n|} \sum_{\mathbf{c} \in \mathcal{C}_n} \sum_{k=1}^K \left(\mathbf{c}^* W_n^* C_k W_n \mathbf{c}\right)^2 + \left(\mathbf{c}^* W_n^* \hat{C}_k W_n \mathbf{c}\right)^2\\
\geq & \frac{K(2K-1)}{2}\sum_{n=1}^N  \frac{|\mathcal{C}_n|}{|\mathcal{C}|}\\
& \cdot \left[\sum_{k=1}^K \operatorname{Tr}\left(C_k W_n \left( \frac{1}{|\mathcal{C}_n|}\sum_{\mathbf{c} \in \mathcal{C}_n}\mathbf{c} \mathbf{c}^*\right)W_n^*\right)^2 + \operatorname{Tr}\left(\hat{C}_k W_n \left(\frac{1}{|\mathcal{C}_n|}\sum_{\mathbf{c} \in \mathcal{C}_n}\mathbf{c} \mathbf{c}^*\right)W_n^* \right)^2 \right]\\
\approx & \frac{K(2K-1)}{2} \sum_{k=1}^K \operatorname{Tr}\left(C_k\right)^2 + \operatorname{Tr}\left(\hat{C}_k\right)^2\\
= & K^2 (2K-1).
\end{split}
\label{eq:lower_bound}
\end{equation}
In the above inequalities, we have used $\operatorname{Tr}(C_k) = \operatorname{Tr}(\hat{C}_k) = 1$, $\sum_{n=1}^N |\mathcal{C}_n| = |\mathcal{C}|$, and $g(\mathcal{C}_n) \approx E$. Note that $(\mathbf{c}^*W_n^*C_kW_n\mathbf{c})^2$ and $(\mathbf{c}^*W_n^*\hat{C}_kW_n\mathbf{c})^2$ are convex with respect to $\mathbf{c}$ since $W_n^*C_kW_n$ and $W_n^*\hat{C}_kW_n$ are positive semidefinite matrices, and the square function is convex and non-decreasing on the non-negative domain (their convexity can be proven in the same way to Theorem 5.1 in \cite{rockafellar}). From the above inequality, there is a tight lower bound of the quantity $R(\{W_n\mathcal{C}_n\})_{n=1}^N$ which is independent of the unitary matrices $W_n$ when each quantity $g(\mathcal{C}_n)$ is the identity matrix. Thus, we conclude that the quantities $g(\mathcal{C}_n)$ should be far from identity. Here, as seen in the way to choose the subsets $\mathcal{C}_n$, we have assumed that $\mathbf{c}$ is randomly and independently chosen and that the average of $\mathbf{c}$ equals zero. Then, by the Law of Large Numbers, the quantity $g(\mathcal{C}_n)$ may be closer to the identity matrix as the number of components in $\mathcal{C}_n$ increases. In such a situation, the lower bound in Eq. (\ref{eq:lower_bound}) may be dominant. For these reasons, since each number of components in the subsets with $N=5$ is larger than one with $N=10$, the converged PAPR with $N=10$ is lower than one with $N=5$.

\begin{figure}[htbp]
\centering  
\includegraphics[width=3.0in]{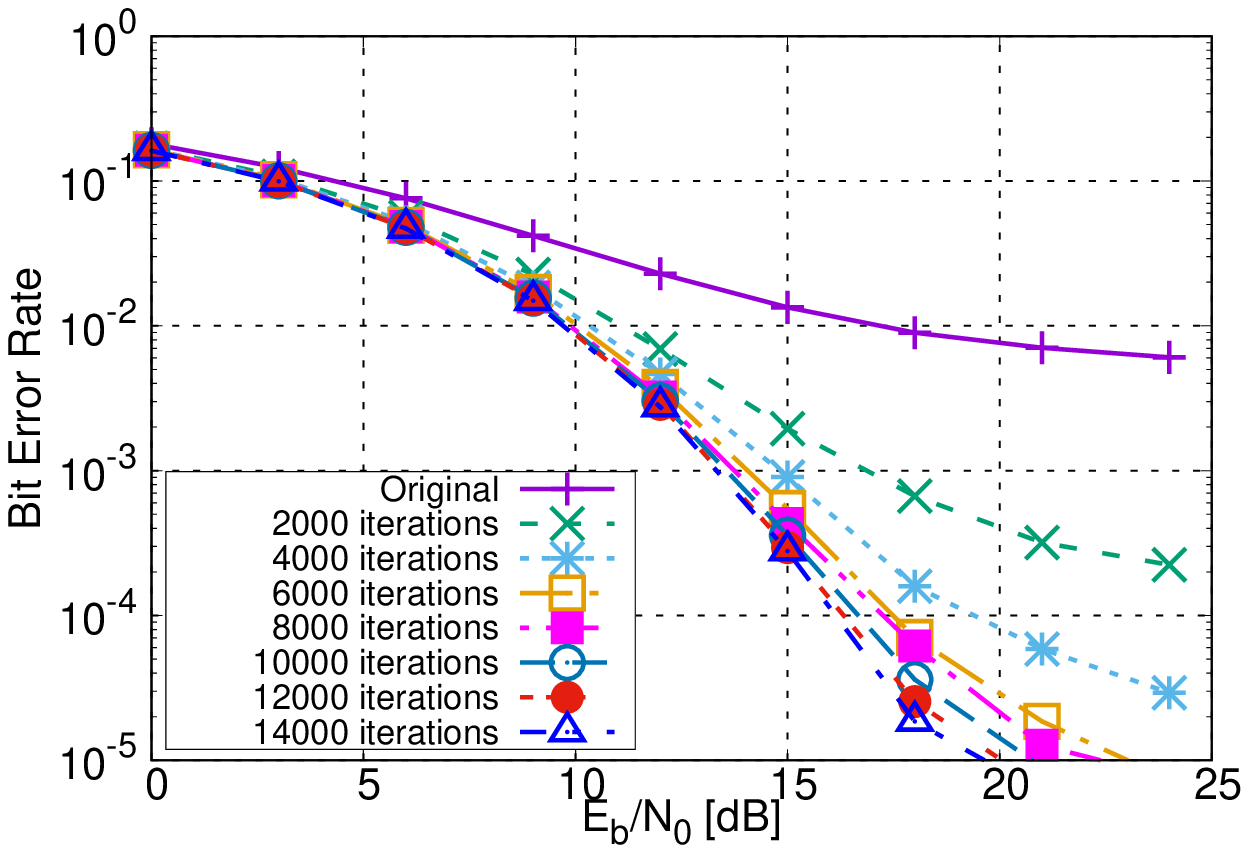}
\caption{Bit Error Rate in each iteration with $N=5$}
\label{fig:BER_iter_N5}
\end{figure}

\begin{figure}[htbp]
\centering  
\includegraphics[width=3.0in]{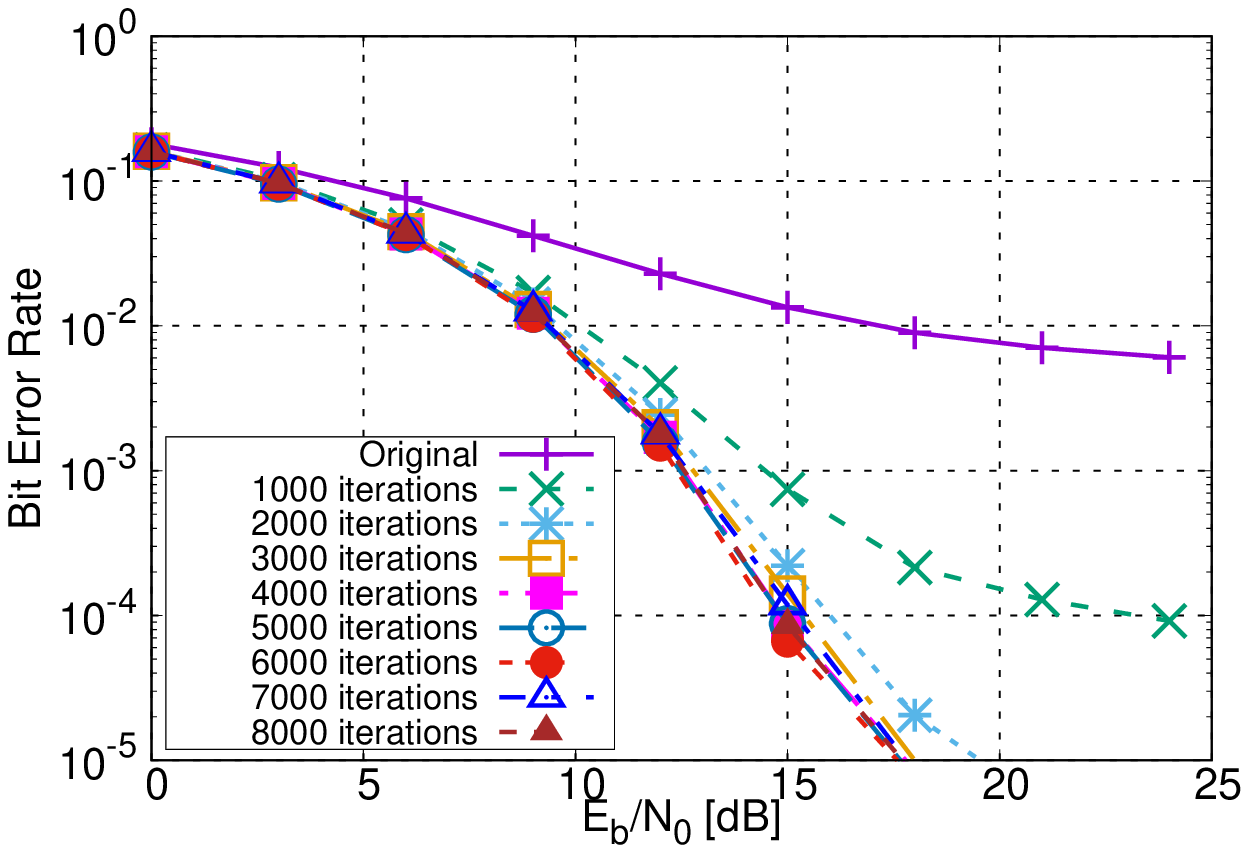}
\caption{Bit Error Rate in each iteration with $N=10$}
\label{fig:BER_iter_N10}
\end{figure}

We have verified that the bound shown in Eq. (\ref{eq:bound_hoeffding}) relates to the CCDF of PAPR and that our method can reduce PAPR. Figures \ref{fig:BER_iter_N5} and \ref{fig:BER_iter_N10} show the BERs with $N=5$ and $N=10$. In these figures, $E_b$ and $N_0$ denote the average energy per bit at the receiver and the variance of the channel Gaussian noise, respectively. The way to choose codewords for each subsets is the same as one in Figs \ref{fig:PAPR_iter_N5} and \ref{fig:PAPR_iter_N10}. To measure the BER, the oversampling parameter $J=1$ has been used \cite{clip}. For an amplifier model, we have used the Rapp model \cite{rapp}, which is described below. Let the input signal be presented in polar coordinates, 
\begin{equation}
 x(t) = \rho(t)\exp(j \theta(t)).
\end{equation}
Then, the output signal is written as 
\begin{equation}
 \zeta(x(t)) = \gamma(\rho(t)) \cdot \exp(j \cdot (\theta(t) + \Phi(\rho(t)))),
\end{equation}
where $\gamma$ and $\Phi$ are functions of the amplitude $\rho(t)$. In the Rapp model, these two functions are chosen as
\begin{equation}
 \gamma(\rho) = \frac{\rho}{\left(1 + \left(\frac{\rho}{r} \right)^{p}\right)^{\frac{1}{2p}}}, \qquad \Phi(\rho) = 0, 
\end{equation}
where $\rho$ is an amplitude, $r$ is the clipping level and $p$ is the real parameter. The parameter $p$ is often chosen as $p=2$ or $p=3$ as seen in \cite{linear_comp}-\cite{rvan}. Here, we have set the parameters $p=2$ and $r = \sqrt{P_{\operatorname{av}}}10^{\frac{1}{10}}$ ($r=2$ [dB]). As seen in these two figures, BER with $N=10$ is lower than one with $N=5$. We have seen that our method with $N=10$ can achieve lower PAPR than one with $N=5$. Thus, we conclude that our method with $N=10$ can achieve lower BER than one with $N=5$ since our method with $N=10$ achieves lower PAPR than one with $N=5$. If the number of codewords is fixed, then it is expected that lower PAPR and BER will be achieved as the number of the subsets gets larger.

\section{Conclusion}
In this paper, we have shown the bound of the CCDF of PAPR and our method to reduce PAPR. The main idea of our method is to transform each subset of codewords with the unitary matrix to reduce the bound of the CCDF of PAPR. Further, the unitary matrices are obtained with the gradient method and the projecting method. In addition, to reduce calculation amount, we have proposed the method with the stochastic gradient method. In the numerical results, the performances of our method have been verified. 

As seen in Section VII, it may not be straightforward to reduce PAPR with our method when the quantity $g(\mathcal{C}_n)$ is nearly
 equivalent to the identity matrix. This obstacle may be overcome when we choose appropriately the subsets of codewords $\mathcal{C}_n$. Therefore, one of remained issues is to explore how to obtain the subsets of codewords $\mathcal{C}_n$. Further, it is necessary to explore other methods to reduce our bound.

\appendices
\section{Proof of Bound with Gaussian Input}

In this appendix, we prove the following relation written in Eq. (\ref{eq:bound_gauss}) 
 \begin{equation*}
  \begin{split}
  &\operatorname{Pr}(\operatorname{PAPR} > \gamma)\\
\leq & \frac{3K(2K-1)}{2P_{\operatorname{av}}^2 \gamma^2} \sum_{k=1}^K \left\{\operatorname{Tr}\left(C_k \Sigma\right)^2 + \operatorname{Tr}\left(\hat{C}_k \Sigma\right)^2\right\}.
  \end{split}
 \end{equation*}
for codewords generated from the complex multivariate Gaussian distribution.

From Eq, (\ref{eq:prob_bound}), it is sufficient to prove
\begin{equation}
 \mathbb{E}\left[\left(\mathbf{c}^* G \mathbf{c}\right)^2\right] 
\leq 3 \operatorname{Tr}\left(G \Sigma\right)^2,
\label{eq:proof_obj}
\end{equation}
where $G$ is a Hermitian and positive semidefinite matrix and $\mathbf{c} \sim \mathcal{CN}(\mathbf{0},\Sigma)$. In \cite{graphical}, it has been shown that $\mathcal{T}(\mathbf{z}) \sim \mathcal{N}\left(\mathcal{T}(\boldsymbol{\mu}),\frac{1}{2}\mathcal{T}(\Sigma)\right)$ if $\mathbf{z} \sim \mathcal{CN}(\boldsymbol{\mu},\Sigma)$, where $\mathcal{N}\left(\boldsymbol{\mu},\Sigma\right)$ denotes the multivariate Gaussian distribution whose mean and covariance matrix are $\boldsymbol{\mu}$ and $\Sigma$, respectively. Here, the transformations $\mathcal{T}(\mathbf{z})$ and $\mathcal{T}(Z)$ are defined as
\begin{equation}
  \mathcal{T}(\mathbf{z}) = \left( \begin{array}{c}
                                     \operatorname{Re}\{\mathbf{z}\}\\
                                     \operatorname{Im}\{\mathbf{z}\}
                                   \end{array} \right), \qquad
\mathcal{T}(Z) = \left( \begin{array}{cc}
                                     \operatorname{Re}\{Z\} & -\operatorname{Im}\{Z\}\\
                                     \operatorname{Im}\{Z\} & \operatorname{Re}\{Z\}
                                   \end{array} \right)
\end{equation}
for $\mathbf{z} \in \mathbb{C}^n$ and $Z \in \mathbb{C}^{n \times n}$.
Note that the matrices $\mathcal{T}(G)$ and $\mathcal{T}(\Sigma)$ are symmetric and positive semidefinite since $G$ and $\Sigma$ are Hermitian and positive semidefinite \cite{telatar}.
From this result, it follows that $\mathcal{T}(\mathbf{c}) \sim \mathcal{N}\left(\mathcal{T}(\mathbf{0}),\frac{1}{2}\mathcal{T}(\Sigma)\right)$. With this and discussions in \cite{telatar}, the left hand side of Eq. (\ref{eq:proof_obj}) is rewritten as
\begin{equation}
\begin{split}
 &\mathbb{E}\left[\left(\mathbf{c}^* G \mathbf{c}\right)^2\right]\\
=  & \mathbb{E}\left[\left(\mathcal{T}(\mathbf{c})^\top \mathcal{T}(G) \mathcal{T}(\mathbf{c})\right)^2\right]\\
= & \sum_{i,j,k,l} \hat{g}_{i,j}\hat{g}_{k,l}\mathbb{E}[\hat{c}_i\hat{c}_j\hat{c}_k\hat{c}_l],
\end{split}
\end{equation}
where $\hat{g}_{i,j}$ and $\hat{c}_k$ are the $(i,j)$-th element of $\mathcal{T}(G)$ and $k$-th element of $\mathcal{T}(\mathbf{c})$, respectively. Note that $\hat{c}_i$ corresponds to either $\operatorname{Re}\{A_i\}$ or $\operatorname{Im}\{A_{i-K}\}$ since each codeword consists of transmitted symbols. In \cite{multivariate}, for $\mathbf{x} \sim \mathcal{N}(\boldsymbol{\mu},\Sigma)$, the fourth moment about the mean has been derived as
\begin{equation}
 \begin{split}
& \mathbb{E}[(x_i - \mu_i)(x_j - \mu_j)(x_k - \mu_k)(x_l - \mu_l)]\\
= & \sigma_{i,j}\sigma_{k,l} + \sigma_{i,k}\sigma_{j,l} + \sigma_{i,l}\sigma_{j,k},
  \end{split}
\label{eq:4thmoment}
\end{equation}
where $x_i$, $\mu_i$ and $\sigma_{i,j}$ are the $i$-th element of $\mathbf{x}$, the $i$-th element of $\boldsymbol{\mu}$ and the $(i,j)$-th element of the real valued-covariance matrix $\Sigma$, respectively. With Eq. (\ref{eq:4thmoment}), Eq. (\ref{eq:proof_obj}) is rewritten as
\begin{equation}
\begin{split}
 &\mathbb{E}\left[\left(\mathbf{c}^* G \mathbf{c}\right)^2\right]\\
= & \frac{1}{4}\left\{\sum_{i,j,k,l} \hat{g}_{i,j}\hat{g}_{k,l}(\hat{\sigma}_{i,j}\hat{\sigma}_{k,l} + \hat{\sigma}_{i,k}\hat{\sigma}_{j,l} + \hat{\sigma}_{i,l}\hat{\sigma}_{j,k})\right\}\\
= & \frac{1}{4}\left\{\operatorname{Tr}(\mathcal{T}(G)\mathcal{T}(\Sigma))^2\right.\\
& + \left. 2\operatorname{Tr}(\mathcal{T}(G)\mathcal{T}(\Sigma)\mathcal{T}(G)\mathcal{T}(\Sigma))\right\},
\end{split}
\label{eq:4thmoment_expand}
\end{equation}
where $\hat{\sigma}_{i,j}$ is the $(i,j)$-th element of $\mathcal{T}(\Sigma)$. In deriving the above second equality, we have used the property that the matrices $\mathcal{T}(G)$ and $\mathcal{T}(\Sigma)$ are symmetric. Let $V$ be a matrix such that $VV^\top = \mathcal{T}(G)$, where such a $V$ can be found since $\mathcal{T}(G)$ is a positive semidefinite. With this decomposition, the relations
\begin{equation}
 \begin{split}
& \operatorname{Tr}(\mathcal{T}(G)\mathcal{T}(\Sigma)\mathcal{T}(G)\mathcal{T}(\Sigma))\\
= & \operatorname{Tr}(VV^\top\mathcal{T}(\Sigma)VV^\top\mathcal{T}(\Sigma))\\
= & \operatorname{Tr}(V^\top\mathcal{T}(\Sigma)V \cdot V^\top\mathcal{T}(\Sigma)V)\\
\leq & \operatorname{Tr}(V^\top\mathcal{T}(\Sigma)V)^2\\ 
= & \operatorname{Tr}(\mathcal{T}(\Sigma)\mathcal{T}(G))^2
\end{split}
\label{eq:4thmoment_right}
\end{equation} 
are obtained.
In the above relations, we have used the properties that the matrix $V^\top\mathcal{T}(\Sigma)V$ is positive semidefinite and $\operatorname{Tr}(XX) \leq \operatorname{Tr}(X)^2$ for any positive semidefinite matrix $X$. In \cite{maxcut}, it has been shown that $\operatorname{Tr}(\mathcal{T}(X)\mathcal{T}(Y)) = 2\operatorname{Tr}(XY)$ for positive semidefinite matrices $X$ and $Y$. Combining this result and Eqs. (\ref{eq:4thmoment_expand}) (\ref{eq:4thmoment_right}), we arrive at the relation
\begin{equation}
\mathbb{E}\left[\left(\mathbf{c}^* G \mathbf{c}\right)^2\right] \leq  3 \operatorname{Tr}\left(G \Sigma\right)^2.
\end{equation}
This is the desired result.

We have proven Eq. (\ref{eq:proof_obj}) for codewords generated from the complex multivariate Gaussian distribution. For codewords generated from the multivariate Gaussian distribution $\mathcal{N}(\mathbf{0},\Sigma)$, we have the same bound in Eq. (\ref{eq:proof_obj}). Thus, Eq. (\ref{eq:bound_gauss}), the bound with Gaussian inputs, is valid for codewords generated from the multivariate Gaussian distribution $\mathcal{N}(\mathbf{0},\Sigma)$.

\section{Proof of Bound with Practical Scheme}
In this appendix, we prove the following bound
 \begin{equation*}
  \begin{split}
 &\operatorname{Pr}(\operatorname{PAPR} > \gamma)\\
\leq & \exp\left(-\frac{2}{(b^2-a^2)}\left(P^2_{\operatorname{av}}\gamma^2 - R\right)^2\right),
  \end{split}
 \end{equation*}
where
\begin{equation*}
R = \frac{K(2K-1)}{2}\sum_{k=1}^K \mathbb{E}\left[\left(\mathbf{c}^* C_k \mathbf{c}\right)^2 + \left(\mathbf{c}^* \hat{C}_k \mathbf{c}\right)^2\right],
\end{equation*}
and the numbers $a$, $b$ and $\gamma$ are given by $a = P_{\operatorname{av}} \cdot \min_{\mathbf{c} \in \mathcal{C}}\operatorname{PAPR}(\mathbf{c}), b = P_{\operatorname{av}} \cdot \max_{\mathbf{c} \in \mathcal{C}}\operatorname{PAPR}(\mathbf{c})$
and $\gamma > R/P_{\operatorname{av}}^2$, respectively.

First, from the Chernoff bound \cite{prob}, we obtain
\begin{equation}
 \operatorname{Pr}(\operatorname{PAPR} > \gamma) \leq \frac{\mathbb{E}\left[\exp(s \cdot \max_t|s(t)|^4)\right]}{\exp(s \cdot P^2_{\operatorname{av}}\gamma^2)},
\end{equation}
where $s > 0$. The parameter $s$ is optimized later. From Hoeffding's Lemma \cite{hoeffding}, it follows
\begin{equation}
\begin{split}
&\frac{\mathbb{E}\left[\exp(s \cdot \max_t|s(t)|^4)\right]}{\exp(s \cdot P^2_{\operatorname{av}}\gamma^2)}\\
 \leq& \frac{\exp\left(s\mathbb{E}\left[|s(t)|^4\right] + (b^2-a^2)s^2/8\right)}{\exp(s \cdot P^2_{\operatorname{av}}\gamma^2)}
\end{split}
\end{equation}
Since $s > 0$ and Eq. (\ref{eq:bound_expansion2}), we have 
\begin{equation}
 \operatorname{Pr}(\operatorname{PAPR} > \gamma) \leq \frac{\exp\left(sR + (b^2-a^2)s^2/8\right)}{\exp(s \cdot P^2_{\operatorname{av}}\gamma^2)}.
\end{equation}
Second, we optimize the parameter $s$. The above inequality is rewritten as
\begin{equation}
\begin{split}
 &\operatorname{Pr}(\operatorname{PAPR} > \gamma)\\
 \leq & \exp\left(-s(P^2_{\operatorname{av}}\gamma^2 - R) + (b^2-a^2)s^2/8\right).
\end{split}
\end{equation}
If $\gamma^2 > R/P_{\operatorname{av}}^2$, then r.h.s of the above inequality has the minimum value at $s=4(P_{\operatorname{av}}^2\gamma^2 - R)/(b^2-a^2)$. Finally, we arrive at
 \begin{equation}
  \begin{split}
 &\operatorname{Pr}(\operatorname{PAPR} > \gamma)\\
\leq & \exp\left(-\frac{2}{(b^2-a^2)}\left(P^2_{\operatorname{av}}\gamma^2 - R\right)^2\right).
  \end{split}
 \end{equation}

\section*{Acknowledgment}

The author would like to thank Dr. Shin-itiro Goto and Dr. Nobuo Yamashita for their helpful advice.

\ifCLASSOPTIONcaptionsoff
  \newpage
\fi

\end{document}